\documentclass[aps, reprint, 
showpacs,showkeys,superscriptaddress,preprintnumbers,nofootinbib]{revtex4-1} 

\usepackage{amsmath,amssymb,bm,mathrsfs}
\usepackage{hyperref}
\usepackage{url}
\usepackage{graphicx}
\usepackage{color}
\usepackage[T1]{fontenc}
\usepackage{newtxtext,newtxmath}

\definecolor{pastelblue}{cmyk}{1,0.34,0,0}
\definecolor{chromegreen}{cmyk}{0.9,0.15,0.8,0}
\definecolor{scarlet}{cmyk}{0,0.8,0.75,0}

\hypersetup{
    colorlinks=true, 
    linkcolor=pastelblue, 
    citecolor=scarlet,
    urlcolor=chromegreen
    }

\begin{document}

\renewcommand{\thesection}{\Roman{section}}


\title{Axionic Wormholes with $R^2$ Correction in Metric and Palatini Formulations} 

\author{Yoshiki Kanazawa}
\email{kanazawa@hep-th.phys.s.u-tokyo.ac.jp}
\affiliation{Department of Physics, University of Tokyo, 
Tokyo 113--0033,
Japan}

\begin{abstract}

QCD axion models have been proposed as a solution to the strong CP problem. QCD instanton effects explicitly violate the global U(1) Peccei-Quinn (PQ) symmetry, and the axion potential is minimized at the CP conserving points. However, it is expected that the global U(1) PQ symmetry is explicitly violated by quantum gravity. This gravitational violation may potentially undermine the PQ solution to the strong CP problem, which is referred to as the axion quality problem.
One source of explicit PQ violation is axionic wormholes, such as the Giddings-Strominger model. Generally speaking, the extent to which the U(1) PQ symmetry is violated depends on models. 
One straightforward extension of the Einstein gravity involves quadratic scalar curvature, which is motivated in cosmological contexts, such as inflation. 
The aim of this work is to investigate the effects of $R^2$ corrections on axionic wormholes. In this paper, we compute wormhole solutions and evaluate the size of explicit PQ violation in the models with the $R^2$ term. Furthermore, We illustrate the differences between the metric and Palatini formulations.

\vspace{0cm}

\end{abstract}

\maketitle

\section{Introduction}

Gravitational violation of global symmetries plays a crucial role in high energy physics. An intriguing example arises in the context of QCD axion models~\cite{Weinberg:1977ma,Wilczek:1977pj}. These models have been proposed as dynamical solutions to the strong CP problem, which involves the fine-tuning issue of CP violation in the QCD sector. Axion is introduced as a Nambu-Goldstone boson associated with the spontaneous breaking of a new global U(1) symmetry, known as the U(1) Peccei-Quinn (PQ) symmetry~\cite{Peccei:1977hh,Peccei:1977ur}. The vacuum expectation value of the axion field determines the effective theta angle. Measurements of the neutron electric dipole moment (EDM) impose a severe constraint on the effective theta angle, $|\theta_{\mathrm{eff}}|\lesssim10^{-10}$~\cite{Abel:2020pzs}. Axions acquire a periodic potential from QCD instantons, and this potential is minimized at CP-conserving points, resulting in the vanishing of the neutron EDM.

However, in the presence of gravity, the PQ solution to the strong CP problem may face challenges. Global symmetries, including the U(1) PQ symmetry, are expected to be explicitly violated by gravity~\cite{Banks:2010zn,Witten:2017hdv,Harlow:2018jwu,Harlow:2018tng}. The axion potential acquires gravitational corrections, which lead to a shift in the vacuum expectation value of the axion field. The PQ solution is particularly sensitive to the corrections due to the stringent constraint on the effective theta angle. To satisfy this constraint, gravitational corrections to the axion potential must be significantly suppressed. This challenge is commonly referred to as the axion quality problem~\cite{Holman:1992us,Kamionkowski:1992mf,Barr:1992qq}.

Axionic wormholes~\cite{Giddings:1987cg,Lee:1988ge,Kallosh:1995hi} are gravitational instantons that feature a configuration connecting two asymptotically flat regions within our universe. The PQ charges pass through these wormholes, resulting in explicit PQ violation within our universe.
At low energy scales, the effects of wormholes are represented as effective local operators that violate the U(1) PQ symmetry~\cite{Giddings:1988cx,Coleman:1988cy,Rey:1989mg,Abbott:1989jw}. These operators experience exponential suppression due to the wormhole action, denoted as $S$.
Recent research on axionic wormholes can be found in Refs.~\cite{Alonso:2017avz,Hebecker:2018ofv,Alvey:2020nyh,Hamaguchi:2021mmt,Cheong:2022ikv}. In the original model~\cite{Giddings:1987cg,Lee:1988ge}, where the axion is the sole matter field and minimally couples to gravity, the wormhole action scales as $S \sim \frac{M_P}{f_a}$, with $f_a$ representing the axion decay constant. PQ violation is sufficiently suppressed when $f_a \lesssim 10^{16}~\mathrm{GeV}$.

Remarkably, the extent of PQ violation varies significantly depending on the specific models employed. When the axion is introduced as the phase component of a complex scalar field, the wormhole action scales as $S\sim\log(\frac{M_P}{f_a})$~\cite{Kallosh:1995hi}. Due to its weak suppression, we encounter the axion quality problem. However, it is possible to enhance the wormhole action by introducing a non-minimal gravitational coupling for the complex scalar field~\cite{Hamaguchi:2021mmt}, thereby achieving sufficient suppression of PQ violation. Moreover, it is worth noting that the physics in extended models of the Einstein gravity can be formulation-dependent. In fact, wormhole solutions in models featuring the non-minimal gravitational coupling differ between the metric and Palatini formulations~\cite{Cheong:2022ikv}. This highlights the influence of gravitational theory extensions on wormhole solutions.

Gravitational theories beyond the Einstein gravity have been recognized for their ability to offer various theoretical advantages. A prominent example is their role in addressing the non-renormalizability issue of 4d Einstein gravity~\cite{Goroff:1985sz,Goroff:1985th}. It has been suggested that quadratic curvatures can render these theories renormalizable~\cite{Stelle:1976gc}. For further insights into quadratic gravity, please refer to Ref.~\cite{Salvio:2018crh}.
Another significant achievement is the realization of cosmological inflation. In the metric formulation, theories involving the square of the scalar curvature imply the existence of an additional dynamical degree of freedom, which is a common feature within the framework of $f(R)$ theories~\cite{Sotiriou:2008rp,Faraoni:2008mf,DeFelice:2010aj}. This additional degree of freedom can represent the inflaton field responsible for cosmological inflation during the early universe~\cite{Starobinsky:1980te}. In the Palatini formulation, inflationary scenarios have been explored in models incorporating the $R^2$ term~\cite{Enckell:2018hmo,Antoniadis:2018ywb,Dimopoulos:2020pas}, even though no such additional degree of freedom exists in this case.
It is worth noting that in the Palatini formulation, the $R^2$ model is equivalent to a model involving higher-order derivatives, which is discussed as an example that implies the axionic weak gravity conjecture~\cite{Andriolo:2020lul}.

In this paper, we investigate axionic wormholes with the $R^2$ correction as a straightforward extension of the Giddings-Strominger model. Even with this simple extension, the formulation of wormhole solutions exhibits significant differences from the Giddings-Strominger model. Here, we provide an overview of this paper. In Section~\ref{sec:R2_model}, we present the formulation of the $R^2$ model for both the metric and Palatini cases. In Sections~\ref{sec:metric} and~\ref{sec:Palatini}, we compute wormhole solutions for the model with the $R^2$ term in both formulations, respectively. In Section~\ref{sec:quality}, we address the axion quality problem within our framework. Finally, in Section~\ref{sec:conc}, we provide our conclusions and engage in a discussion of our findings.

\section{$R^2$ Model}
\label{sec:R2_model}

We study the Euclidean axionic wormholes in the models with $R^2$ term. This model is included in the framework of $f(R)$ theories~\cite{Sotiriou:2008rp,Faraoni:2008mf,DeFelice:2010aj},
\begin{align}
    f(R) = R + \frac{\eta}{2} \frac{R^2}{M_P^{2}},
\end{align}
where $M_P=(8 \pi G)^{-1/2}\simeq 2.44 \times 10^{18}~\mathrm{GeV}$ represents the reduced Planck mass, and $\eta$ is a dimensionless constant. In this study, we consider only the axion as the matter degree of freedom. The action in the Jordan frame is given by
\begin{align}
    S&=\int d^4x~\sqrt{g_J}\left[
    -\frac{M_P^2}{2} f(R_J) + \frac{f_a^2}{2}(\partial_{\mu}\theta)^2
    \right],
    \label{eq:R2_action}
\end{align}
where $g_J$ is the determinant of the metric tensor in the Jordan frame and the axion field $\theta$ exhibits $2\pi$-periodicity, i.e., $\theta\sim\theta + 2\pi$.
By introducing an auxiliary field, $\phi$, we have an equivalent action,
\begin{align}
    S
    &=\int d^4x~\sqrt{g_J}\left[
    -\frac{M_P^2}{2} (f^{\prime}(\phi) (R_J - \phi) + f(\phi)) 
    \right.
    \nonumber \\
    &\quad\quad \left.
    + \frac{f_a^2}{2}(\partial_{\mu}\theta)^2
    \right]
    \\
    &=\int d^4x~\sqrt{g_J}\left[
    -\frac{M_P^2}{2} \Omega^2(\phi) R_J + \tilde{U}(\phi) + \frac{f_a^2}{2}(\partial_{\mu}\theta)^2
    \right].
    \nonumber
\end{align}
Here,
\begin{align}
    \Omega^2(\phi)&\equiv f^{\prime}(\phi) 
    = 1 + \eta\frac{\phi}{M_P^{2}},
    \\
    \tilde{U}(\phi)&\equiv -\frac{M_P^2}{2} [f(\phi) - \phi f^{\prime}(\phi)]
    = \eta   \frac{\phi^{2}}{4}.
\end{align}
Variation with respect to $\phi$ yields
\begin{align}
    f^{\prime \prime}(\phi)(R_J - \phi) = 0,
\end{align}
resulting in the reproduction of Eq. \eqref{eq:R2_action}. 

We perform the Weyl transformation,
\begin{align}
    g_{\mu\nu}=\Omega^2 g_{J\mu\nu}.
    \label{eq:Weyl_transformation}
\end{align}
The transformation law of the Ricci scalar depends on the formulation of gravity. We discuss two formulations; the metric and Palatini cases.

\subsection{Metric formulation}

In the metric farmulation, the affine connection is introduced as the Levi-Civita connection. The Ricci scalar transforms as 
\begin{align}
    R_J=\Omega^2\left[
    R + 3 \nabla^2 \log\Omega^2 -\frac{3}{2}g^{\mu\nu}\partial_{\mu}\log\Omega^2 \partial_{\nu}\log\Omega^2
    \right],
\end{align}
where $\nabla^2$ is the d'Alembert operator.
The second term does not affect the equations of motion, because it has the form of a total derivative. Moreover, it has no contribution to the value of the action under boundary conditions we consider later. Therefore, we ignore this term. The Einstein frame action is given by
\begin{align}
    S&=\int d^4x~\sqrt{g}\left[-\frac
    {M_P^2}{2} R + \frac{3M_P^2}{4}(\partial_{\mu}\log\Omega^2)^2 
    \right.
    \nonumber \\
    &\quad\quad \left. + U(\phi) + \frac{f_a^2}{2\Omega^2}(\partial_{\mu}\theta)^2 \right],
\end{align}
where
\begin{align}
    U(\phi)&\equiv \frac{\tilde{U}(\phi)}{\Omega^4(\phi)}.
    \label{eq:pot}
\end{align}
The presence of the kinetic term of $\phi$ implies that $\phi$ acts as a dynamical degree of freedom. To avoid the tachyonic instability, we choose $\eta>0$.


\subsection{Palatini formulation}

In the Palatini formulation, the affine connection $\Gamma^{\lambda}_{\mu\nu}$ and the metric $g_{\mu\nu}$ are treated as independent quantities in the action, and the Ricci tensor $R_{\mu\nu} = R_{\mu\nu}(\Gamma,\partial\Gamma)$ is introduced as an explicit function with respect to the affine connection. Under the Weyl transformation, the Ricci scalar $R=g^{\mu\nu}R_{\mu\nu}(\Gamma,\partial\Gamma)$ transforms as 
\begin{align}
    R_J=\Omega^2 R.
\end{align}
The Einstein frame action is given by
\begin{align}
    S&=\int d^4x~\sqrt{g}\left[
    -\frac{M_P^2}{2}R + U(\phi) + \frac{f_a^2}{2\Omega^2}(\partial_{\mu}\theta)^2
    \right].
\end{align}
The notable distinction from the metric case is the absence of any additional degrees of freedom. The variation with respect to $\phi$ results in a constraint on the fields. 
Integrating out $\phi$, we obtain an equivalent action that includes the higher-order derivative,
\begin{align}
    S&=\int d^4x~\sqrt{g}\left[
    -\frac{M_P^2}{2}R + \frac{f_a^2}{2}(\partial_{\mu}\theta)^2 - \frac{\eta}{4}\frac{f_a^4}{M_P^4} [(\partial_{\mu} \theta)^2]^2
    \right].
    \label{eq:higher_deri}
\end{align}
The positivity bound for $\theta\theta\rightarrow\theta\theta$ scattering in the Lorentzian signature indicates $\eta > 0$\footnote{ Ref.~\cite{Andriolo:2020lul} indicates that this positivity bound is demonstrated only when gravity can be ignored. In the presence of gravity, this condition may hold true for $|\eta|$ greater than a lower bound determined by the string scale.}~\cite{Adams:2006sv}.

\section{Analysis in Metric formulation}
\label{sec:metric}

\subsection{Wormhole solutions}

We compute the wormhole solutions in the $R^2$ model for the metric formulation. As we discussed in the previous section, we consider the case of $\eta>0$. We assume the $O(4)$-symmetric geometry, described by $ds^2 = dr^2 + a(r)^2 d\Omega_3^2$, where $r$ is the Euclidean time, and $a$ is the scale factor. We introduce a canonically normalized field, referred to as the scalaron,
\begin{align}
    \omega\equiv \sqrt{\frac{3}{2}} M_P \log\Omega^2.
\end{align}
The Einstein frame action is rewritten as
\begin{align}
    S&=\int d^4x~\sqrt{g}\left[-\frac
    {M_P^2}{2} R + \frac{1}{2}(\partial_{\mu}\omega)^2 + U(\omega)
    \right.
    \nonumber \\
    &\quad\quad \left.+ e^{-\sqrt{\frac{2}{3}} \frac{\omega}{M_P}} \frac{f_a^2}{2}(\partial_{\mu}\theta)^2\right],
\end{align}
where the potential is
\begin{align}
    U(\omega)&=
    \frac{M_P^4}{4 \eta} \left(1 - 
    e^{-\sqrt{\frac{2}{3}}\frac{\omega}{M_P}}
    \right)^2.
    \label{eq:potential}
\end{align}
The potential becomes constant for positive large field values. The scalaron field can serve as the inflaton field to achieve slow-roll inflation, as proposed by Starobinsky~\cite{Starobinsky:1980te}. If the scalaron field assumes the role of the inflaton, we typically find $\eta \sim 10^{9}$ with the scalaron mass of $m_{\omega}\sim10^{13}~\mathrm{GeV}$~\cite{Planck:2018jri}. Even if we choose smaller values of $\eta$, it is possible to realize multi-field inflation, such as the Higgs-scalaron mixed inflation~\cite{Ema:2017rqn}. Strictly speaking, in such models, wormhole solutions may exhibit minor differences from those in the $R^2$ model.

In the metric formulation, the action has the second-order derivative in the Ricci scalar term. To make the variational principle well-defined, we add the Gibbons-Hawking-York (GHY) boundary term~\cite{Gibbons:1976ue,York:1972sj},
\begin{align}
    S_{\mathrm{GHY}}&= -M_P^2 \int_{\partial V} d^3x \sqrt{\tilde{g}} (K - K_0),
\end{align}
where $\tilde{g}$ is the determinant of the induced metric on the boundary $\partial V$, $K$ is the trace of the extrinsic curvature tensor for the boundary and $K_0$ is that for the same boundary embedded into the flat spacetime. The introduction of $K_0$ helps cancel out the divergence of $K$ as $r\rightarrow\infty$. The GHY term does not contribute to the equations of motion.

\begin{figure}[t]
    \centering
    {\includegraphics[width=0.45\textwidth]{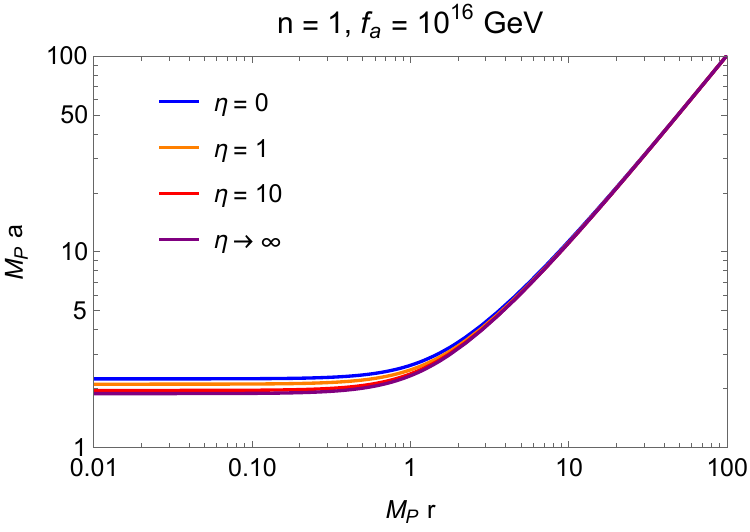}}
    {\includegraphics[width=0.45\textwidth]{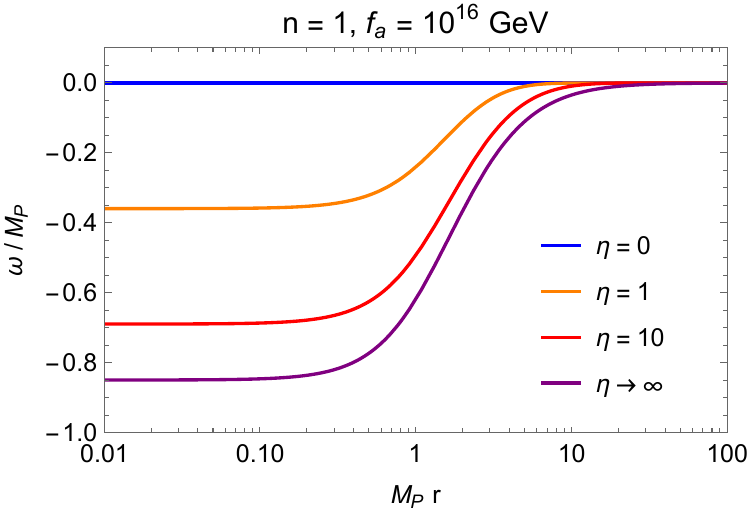}}
    \caption{$a(r)$ and $\omega(r)$ in units of $M_P$ for several values of $\eta$ in the metric formulation.}
    \label{fig:a_and_omega}
\end{figure}

When searching for correct saddle-point solutions by computing the Euclidean path integral, the charge conservation associated with the axion shift symmetry,
\begin{align}
    \partial_{\mu}J^{\mu} = 0,\quad J^{\mu} = g^{\mu\nu}\sqrt{g} f_a^2 e^{-\sqrt{\frac{2}{3}} \frac{\omega}{M_P}} \partial_{\nu} \theta,
    \label{eq:charge_cons}
\end{align}
must be considered as a constraint. To do this, introducing $\theta$ as a Lagrange multiplier and $J^{\mu}$ as a variable of the path integral~\cite{Coleman:1989zu}, we search for saddle-point solutions that minimize the following action,
\begin{align}
    S&=\int d^4x~\sqrt{g}\left[-\frac
    {M_P^2}{2} R + \frac{1}{2}(\partial_{\mu}\omega)^2 + U(\omega)
    \right.
    \nonumber \\
    &\quad\quad \left.+ \frac{e^{\sqrt{\frac{2}{3}} \frac{\omega}{M_P}}}{2 g f_a^2 } J_{\mu} J^{\mu}
    + \frac{1}{\sqrt{g}}\theta \partial_{\mu}J^{\mu}
    \right].
\end{align}
The variations with respect to $\theta$ and $J^{\mu}$ reproduce Eq.~\eqref{eq:charge_cons}.
The conserved charge is quantized by an integer $n$ due to the periodicity of the axion field,
\begin{align}
    2\pi^2 a^3 f_a^2 e^{-\sqrt{\frac{2}{3}}\frac{\omega}{M_P}} \dot{\theta} = n,
\end{align}
where the dot denotes the derivative with respect to $r$. 
From the variations with respect to the metric and $\omega$, we find equations of motion,
\begin{align}
    &\dot{a}^2 - 1
    = \frac{a^2}{3 M_P^2}\left[
    \frac{1}{2}\dot{\omega}^2 - e^{-\sqrt{\frac{2}{3}}\frac{\omega}{M_P}} \frac{f_a^2}{2} \dot{\theta}^2 - U(\omega)
    \right],
    \\
    &a\ddot{a} + \dot{a}^2 - 1
    = \frac{a^2}{3 M_P^2}\left[
    - \frac{1}{2}\dot{\omega}^2 + e^{-\sqrt{\frac{2}{3}}\frac{\omega}{M_P}} \frac{f_a^2}{2} \dot{\theta}^2 - 2U(\omega)
    \right],
    \\
    &\ddot{\omega} + 3 \frac{\dot{a}}{a} \dot{\omega}
    = \frac{dU}{d\omega} + \sqrt{\frac{2}{3}}\frac{1}{M_P} e^{-\sqrt{\frac{2}{3}}\frac{\omega}{M_P}} \frac{f_a^2}{2} \dot{\theta}^2.
\end{align}
We search for asymptotically flat solutions that satisfy the following boundary conditions,
\begin{align}
    \dot{a}(0) = 0,\quad\dot{\omega}(0) = 0,\quad \omega(\infty) = 0.
\end{align}
The wormhole solutions continuously vary from $\eta=0$ to $\eta\rightarrow\infty$, as shown in Fig. \ref{fig:a_and_omega}. The size of the wormhole throat, $a(0)$, hardly depends on the parameter $\eta$. 
For large $\eta$, the values at $r=0$ approach certain values. To demonstrate this, we will derive the analytic form in the limit $\eta\rightarrow\infty$ later.

When $\eta$ is large, the inverse size of the wormhole throat $\sim\sqrt{f_aM_P}$ can exceed the UV cutoff scale. The perturbative unitarity cutoff is estimated using power counting analysis~\cite{Burgess:2009ea},
\begin{align}
    \Lambda \sim \begin{cases}
        M_P & \eta \ll 1
        \\
        \frac{M_P}{\eta^{1/3}} & \eta \gg 1
    \end{cases}.
\end{align}
The validity of the effective theory requires that the inverse size of the wormhole throat does not exceed the UV cutoff scale~\cite{Hebecker:2018ofv}. This results in an upper bound of $\eta$ for a given $f_a$.

\subsection{Wormhole action}

\begin{figure}[t]
    \centering
    {\includegraphics[width=0.45\textwidth]{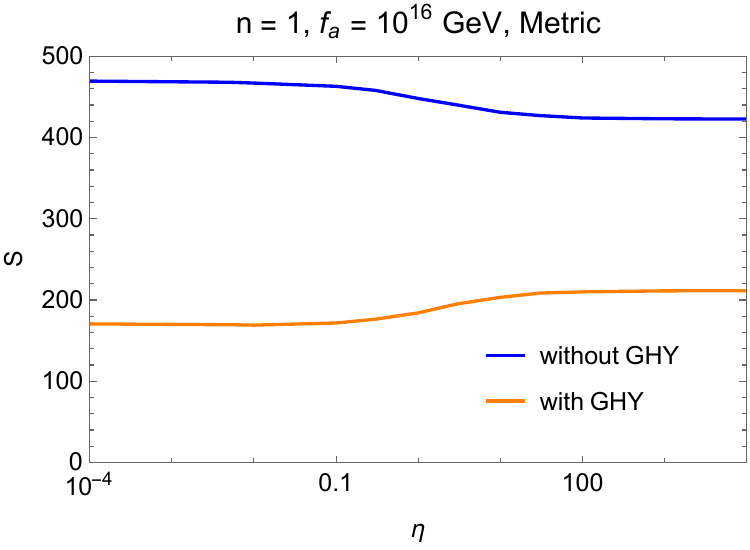}}
    \caption{Wormhole action without and with the Gibbons-Hawking-York boundary term as a function of $\eta$.}
    \label{fig:wormhole_action_metric_R2}
\end{figure}

The total action is composed of two terms; $S_{\mathrm{total}} = S + S_{\mathrm{GHY}}$.
Using the equations of motion, $S$ is given by 
\begin{align}
    S&= 2\pi^2 \int_0^{\infty} dr~\left[
    3M_P^2 a^2 \ddot{a} + a^3 \dot{\omega}^{2}
    \right].
\end{align}
The contribution of the GHY term is
\begin{align}
    S_{\mathrm{GHY}}&= -6\pi^2 M_P^2 a(0)^2,
\end{align}
which takes negative values. The wormhole action without and with the Gibbons-Hawking-York boundary terms are shown in Fig.~\ref{fig:wormhole_action_metric_R2}. The wormhole action hardly depends on $\eta$ and take the value of $\sim \frac{|n| M_P}{f_a}$ for any $\eta$.

\subsection{Analytic formula for $\eta = 0$}

The $\eta = 0$ case corresponds to the Giddings-Strominger model. The wormhole solution is given by the elliptic integrals~\cite{Giddings:1987cg}. The size of the wormhole throat is
\begin{align}
    a(0)=\left(
    \frac{n^2}{24\pi^4 f_a^2 M_P^2}
    \right)^{\frac{1}{4}}.
\end{align}
We can find the analytic formula for the wormhole actions,
\begin{align}
    S=\sqrt{\frac{3\pi^2}{8}}\frac{|n| M_P}{f_a},
\end{align}
and
\begin{align}
    S_{\mathrm{total}}=S+S_{\mathrm{GHY}}=
    \left(1-\frac{2}{\pi}\right) \sqrt{\frac{3\pi^2}{8}}\frac{|n| M_P}{f_a}.
\end{align}

\subsection{Analytic formula for $\eta\rightarrow\infty$}

In the limit of $\eta\rightarrow\infty$, we can analytically solve the equations and calculate the wormhole action.
To provide clarity, we discuss this in the Jordan frame, where the line element is given by $ds^2 = dt^2 + a_J(t)^2 d\Omega_3^2$. In this frame, the quantities are related to those in the Einstein frame as follows; $t = \int dr~\Omega^{-1}$ and $a_J = \Omega^{-1} a$.
We assume that $r=0$ corresponds to $t=0$. 
In the limit $\eta\rightarrow\infty$, the solution remains constant with $\phi = 0$, as given by $\phi = M_P^2 (\Omega^2 - 1) / \eta$.
The relation $\phi = R_J$ implies a flat geometry, $R_J = 0$. Under the boundary condition $da_J/dt|_{t=0} = 0$, we find the hyperbolic solution,
\begin{align}
    a_J(t) = \sqrt{t^2 + a_J(0)^2}.
\end{align}
$a_J(0)$ has not been determined yet. The first-order Einstein equation leads to the equation for $\Omega^2$,
\begin{align}
    \Omega^2 \left[ 
    \left(\frac{d a_J}{dt}\right)^2 -1\right] + a_J \frac{d a_J}{dt} \frac{d \Omega^2}{dt}
    = -\frac{n^2}{24 \pi^4 f_a^2 M_P^2 a_J^4}.
\end{align}
Under the boundary conditions $\Omega^2|_{t\rightarrow\infty} = 1$ and $d\Omega^2 / dt |_{t=0} = 0$, we find
\begin{align}
    a_J(0)&=\left(\frac{n^2}{12\pi^4 f_a^2 M_P^2}\right)^{\frac{1}{4}},
    \\
    \Omega^2(\phi(t))&=\frac{t^2+a_J(0)^2/2}{t^2+a_J(0)^2}.
    \label{eq:sol_xi2=infty}
\end{align}
From the analytic formula, we can calculate the wormhole actions,
\begin{align}
    S&=\frac{\sqrt{3} |n| M_P}{f_a},
\end{align}
and
\begin{align}
    S_{\mathrm{total}}=S+S_{\mathrm{GHY}}=
    \frac{\sqrt{3} |n| M_P}{2f_a},
\end{align}
which agree with the numerical results shown in Fig.~\ref{fig:wormhole_action_metric_R2}.

\section{Analysis in Palatini formulation}
\label{sec:Palatini}

\subsection{Wormhole solutions}

We compute the wormhole solutions in the $R^2$ model for the Palatini formulation. As we discussed in Sec.~\ref{sec:R2_model}, we consider the case of $\eta>0$.
We assume the $O(4)$-symmetric geometry, $ds^2 = dr^2 + a (r) ^2 d\Omega_3^2$, as is in the metric case.
The Einstein frame action is given by
\begin{align}
    S&=\int d^4x~\sqrt{g}\left[
    -\frac{M_P^2}{2}g^{\mu\nu}R_{\mu\nu}(\Gamma,\partial\Gamma) + U(\phi)
    \right.
    \nonumber \\
    &\quad\quad \left.+ \frac{f_a^2}{2\Omega^2}(\partial_{\mu}\theta)^2
    \right],
\end{align}
where the potential is defined as Eq.~\eqref{eq:pot}
\begin{align}
    U(\phi)= \frac{\eta}{4}\frac{\phi^2}{(1 + \frac{\eta}{M_P^2}\phi)^2}.
\end{align}
The $\phi$ field is not a dynamical degree of freedom unlike the metric case. 

As in the metric case, we introduce $\theta$ as a Lagrange multiplier and $J^{\mu}$ as a variable of the path integral, and search for saddle-point solutions that minimize the following action,
\begin{align}
    S&=\int d^4x~\sqrt{g}\left[
    -\frac{M_P^2}{2}g^{\mu\nu}R_{\mu\nu}(\Gamma,\partial\Gamma) + U(\phi)
    \right.
    \nonumber \\
    &\quad\quad \left.+ \frac{\Omega^2}{2 g f_a^2} J_{\mu} J^{\mu}
    + \frac{1}{\sqrt{g}} \theta \partial_{\mu} J^{\mu}
    \right].
    \label{eq:action_Pala}
\end{align}
The variation with respect to $\phi$ leads to,
\begin{align}
    0&=\frac{d}{d\phi}\left[
    \frac{\Omega^2}{2 g f_a^2} J_{\mu} J^{\mu} + U(\phi)
    \right]
    \nonumber \\
    &=  \frac{f^{\prime \prime}(\phi)}{2}
    \left[  \frac{1}{g f_a^2} J_{\mu} J^{\mu}
    + M_P^2 \frac{2f(\phi) - \phi f^{\prime} (\phi)}{f^{\prime 3} (\phi)}
    \right]
    \nonumber \\
    &=  \frac{f^{\prime \prime}(\phi)}{2} \left[
      \frac{1}{g f_a^2} J_{\mu} J^{\mu} + M_P^2 \frac{\phi}{\Omega^6}
    \right]. 
    \label{eq:phi_eom}
\end{align} 
Note that $f^{\prime \prime}(\phi) \neq 0$ in this model.
The variations with respect to $\theta$ and $J^{\mu}$ lead to the charge conservation,
\begin{align}
    \partial_{\mu}J^{\mu} = 0,\quad J^{\mu} = g^{\mu\nu}\sqrt{g} \frac{f_a^2}{\Omega^2} \partial_{\nu} \theta.
    \label{eq:charge_cons2}
\end{align}
The conserved charge is quantized by an integer $n$ due to the periodicity of the axion field,
\begin{align}
    2\pi^2 a^3 \frac{f_a^2}{\Omega^2} \dot{\theta} = n.
    \label{eq:PQ_charge}
\end{align}
Using Eqs.~\eqref{eq:phi_eom},~\eqref{eq:charge_cons2} and~\eqref{eq:PQ_charge}, we obtain a constraint on $a$ and $\phi$,
\begin{align}
     &a^2 = - \left(\frac{n}{2\pi^2 f_a M_P}\right)^{2/3} \frac{\Omega^2}{\phi^{1/3}}.
     \label{eq:a-phi}
\end{align}
This relation implies that $\phi$ takes negative values.
The variation with respect to the affine connection leads to the Levi-Civita connection,
\begin{align}
    \Gamma^{\lambda}_{\mu\nu} = \frac{1}{2}g^{\lambda\alpha}(\partial_{\mu} g_{\alpha\nu} + \partial_{\nu} g_{\mu\alpha} - \partial_{\alpha} g_{\mu\nu} ).
\end{align}
From the variation with respect to the metric, we find
\begin{align}
    &\dot{a}^2 - 1 = \frac{a^2}{3 M_P^2}\left[-\frac{f_a^2}{2\Omega^2} \dot{\theta}^2 - U(\phi)\right],
    \label{eq:Ein_eq-1}
    \\
    &a\ddot{a} + \dot{a}^2 -1 = \frac{a^2}{3 M_P^2}\left[\frac{f_a^2}{2\Omega^2} \dot{\theta}^2 - 2U(\phi)\right],
    \label{eq:Ein_eq-2}
\end{align}
Due to the constraint, these equations have one degree of freedom.

\begin{figure}[t]
    \centering
    {\includegraphics[width=0.45\textwidth]{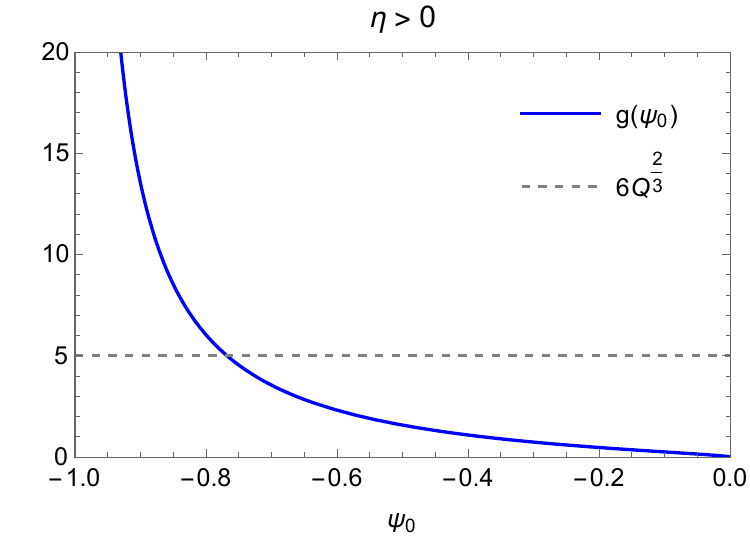}}
    \caption{The initial value $\psi_0\equiv \psi(0)$ is determined as the value at the intersection for a given $Q$.}
    \label{fig:initial}
\end{figure}

\begin{figure}[t]
    \centering
    {\includegraphics[width=0.45\textwidth]{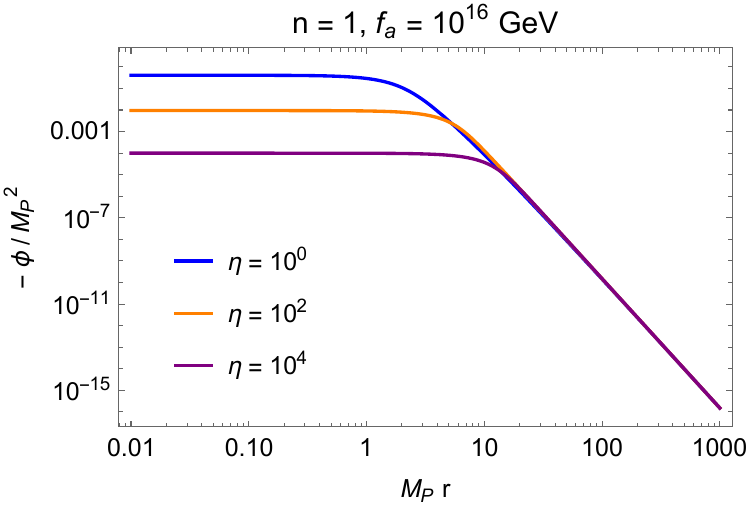}}
    {\includegraphics[width=0.45\textwidth]{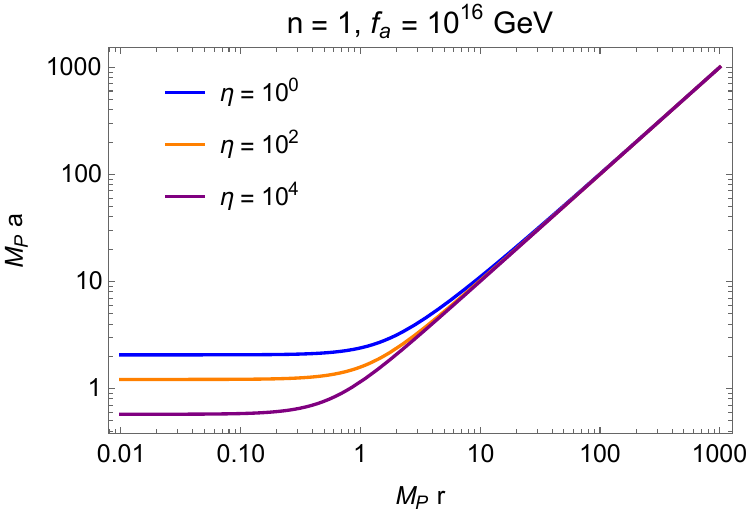}}
    \caption{$\phi(r)$ and $a(r)$ in units of $M_P$ for several values of $\eta$ in the Palatini formulation.}
    \label{fig:psi_and_A}
\end{figure}

We search for asymptotically flat solutions that satisfy the following boundary conditions,
\begin{align}
    \dot{a}(0) = 0 \quad \mathrm{or} \quad\dot{\phi}(0) = 0.
\end{align}
Introducing dimensionless quantities for convenience,
\begin{align}
    \rho=\frac{M_P}{\sqrt{\eta}}r,
    \quad\psi=\frac{\eta}{M_P^2}\phi,
    \quad A=\frac{M_P}{\sqrt{\eta}}a,
    \label{eq:new_quantities}
\end{align}
we rewrite Eqs.~\eqref{eq:Ein_eq-1} and \eqref{eq:Ein_eq-2} as equations for $\psi$,
\begin{align}
    & -\frac{(2\psi -1)^2}{6\psi^{3} (1 + \psi)} \left(\frac{d \psi}{d\rho}\right)^2 - \frac{6Q^{2/3}}{\psi^{2/3}}
    = -\frac{1 - \frac{1}{2} \psi}{1 + \psi},
    \label{eq:eom_R2_P-2}
    \\
    &- \frac{2\psi - 1}{\psi^2} \frac{d^2\psi}{d\rho^2} - \frac{2}{3} \frac{2 - \psi}{\psi^3} \left(\frac{d \psi}{d\rho}\right)^2
    - \frac{6Q^{2/3}}{\psi^{2/3}} 
    = 1,
    \label{eq:eom_R2_P-3}
\end{align}
where $Q$ is defined as
\begin{align}
    Q= \frac{2\pi^2 f_a \eta}{|n| M_P}.
    \label{eq:def_Q}
\end{align}
Substituting the boundary condition $d \psi / d \rho = 0$ at $\rho = 0$ for Eq.~\eqref{eq:eom_R2_P-2}, we find the equation that determines the initial value $\psi(0)$,
\begin{align}
    6Q^{2/3} = \psi(0)^{2/3}\frac{1 - \frac{1}{2} \psi(0)}{1 + \psi(0)}\equiv g(\psi(0)),
    \label{eq:eq_psi0}
\end{align}
whose root is schematically shown in Fig.~\ref{fig:initial}. We have to note that $\psi$ is negative when $\eta>0$. The function $g(\psi_0)$ diverges at $\psi_0\rightarrow-1$. Therefore, for any positive value of $\eta$, the initial value can be determined, and the wormhole solutions exist, as shown in Fig.~\ref{fig:psi_and_A}. 

Using $\Omega^2(\psi(0))\sim Q^{-2/3}$ for large $Q$, we find an asymptotic behavior,
\begin{align}
    \frac{a(0)}{a_{\eta=0}(0)} \sim Q^{-1/6}.
    \label{eq:a_scale}
\end{align}
This implies that the inverse of $a(0)$ can increase infinitely as $\eta$ becomes large. The perturbative unitarity cutoff is estimated using power counting analysis~\cite{Burgess:2009ea},
\begin{align}
    \Lambda\sim\begin{cases}
        M_P & \eta \ll 1
        \\
        \frac{M_P}{\eta^{1/4}} & \eta \gg 1
    \end{cases}.
\end{align} 
The validity of the effective theory~\cite{Hebecker:2018ofv} leads to an upper bound of $\eta$ for a given $f_a$, as in the metric case.




\subsection{Wormhole action}

\begin{figure}[t]
    \centering
    {\includegraphics[width=0.45\textwidth]{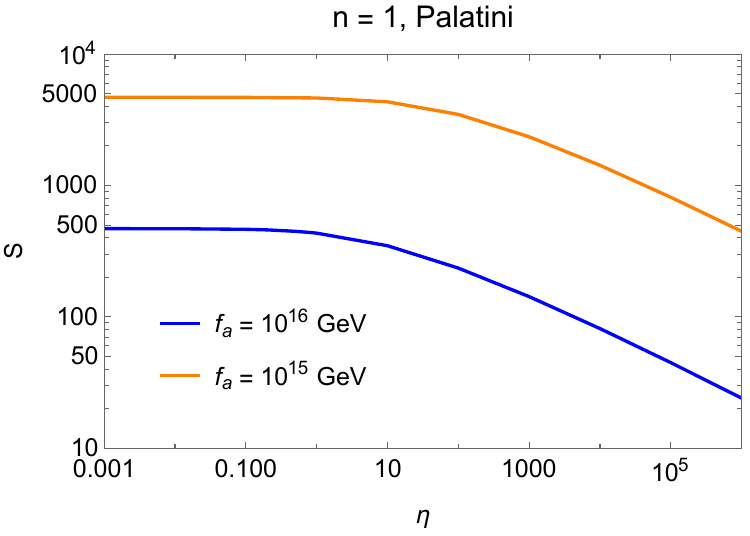}}
    \caption{Wormhole action as a function of $\eta$ in the Palatini formulation.}
    \label{fig:wormhole_action_R2}
\end{figure}

The wormhole action\footnote{In the Palatini formulation, the GHY term is not needed, because the action does not have second-order derivatives.} is given by an integral with respect to $\psi$,
\begin{align}
    S&=2\pi^2\int_0^{\infty} dr~a^3\left[\frac{f_a^2 \dot{\theta}^{2}}{\Omega^2} - U\right]
    \nonumber \\
    &=\frac{|n| M_P}{f_a}\int_0^{\infty}d\rho~\frac{ \sqrt{- \psi} (1 + \frac{1}{4} \psi)}{\sqrt{1+\psi}} \label{eq:wormhole_action-Palatini} \\
    &=\frac{|n| M_P}{f_a}\int_0^{\psi(0)}d\psi~\frac{(1 + \frac{1}{4} \psi) (1- 2\psi)}{\sqrt{6} \psi (1 + \psi)} 
    \left[\frac{6Q^{2/3}}{\psi^{2/3}} - \frac{ 1 - \frac{1}{2} \psi}{1 + \psi}\right]^{-\frac{1}{2}}.
    \nonumber
\end{align}
To obtain the last line, we use the equation of motion~\eqref{eq:eom_R2_P-2}. The integral part is determined by the parameter $Q$. The wormhole action as a function of $\eta$ is shown in Fig. \ref{fig:wormhole_action_R2}. 
The wormhole action exhibits a decreasing trend with respect to $\eta$, which is consistent with the analysis of higher-order derivative corrections to the Giddings-Strominger model~\cite{Andriolo:2020lul}. For large values of $\eta$, the wormhole action experiences significant suppression.

\begin{figure}[t]
    \centering
    {\includegraphics[width=0.45\textwidth]{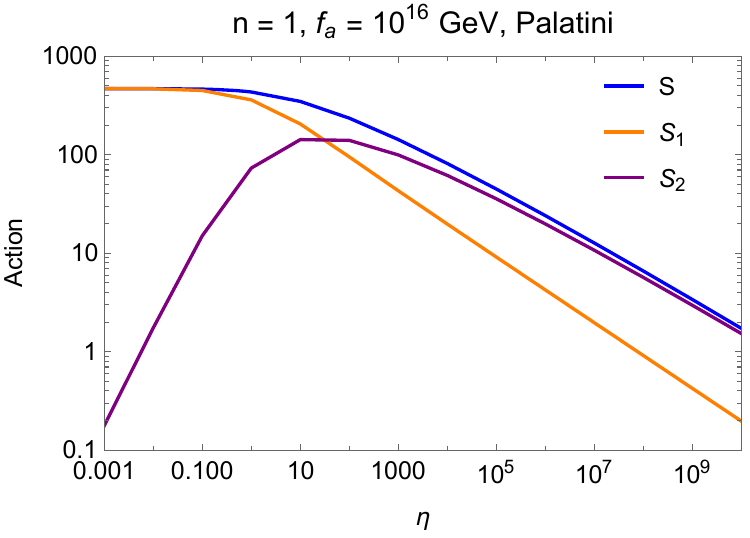}}
    \caption{Wormhole actions $S,S_1,S_2$ as functions of $\eta$.}
    \label{fig:action16}
\end{figure}

To gain a deeper understanding of the wormhole action, using the equations of motion, we represent the action \eqref{eq:action_Pala} as two distinct components, denoted as $S=S_1+S_2$, which are defined as
\begin{align}
    S_1&=\int d^4x~\sqrt{g}\left[
    -\frac{M_P^2}{2}R + \frac{f_a^2}{2}(\partial_{\mu}\theta)^2
    \right] ,
    \\
    S_2&=\int d^4x~\sqrt{g} \frac{3\eta}{4}\frac{f_a^4}{M_P^4} [(\partial_{\mu} \theta)^2]^2.
\end{align}
Fig.~\ref{fig:action16} illustrates that $S_1$ dominates the total action for small $\eta$, and its value is comparable to that of the Giddings-Strominger model. Conversely, for large $\eta$, the higher-order derivative term $S_2$ takes precedence, causing the total action to monotonically decrease, eventually approaching zero.

\section{Axion quality problem}
\label{sec:quality}

In low-energy theories, explicit PQ violation caused by wormholes is represented as effective local operators~\cite{Giddings:1988cx,Coleman:1988cy,Rey:1989mg,Abbott:1989jw}. The dominant contribution to the axion potential arises from the term associated with the unit PQ charge~\cite{Alonso:2017avz,Hebecker:2018ofv},
\begin{align}
    V_{\mathrm{wh}} &= cL^{-4} e^{-S} \cos\left(
    \theta + \delta
    \right),
\end{align}
where $L \equiv a(0)$ is the size of the wormhole throat, and $S$ is the wormhole action. The dimensionless constant $c$ and the relative phase to the QCD contribution $\delta$ cannot be determined without an underlying principle. The effective theta angle is estimated as,
\begin{align}
    \theta_{\mathrm{eff}} &= c\sin\delta \frac{L^{-4}}{\Lambda_{\mathrm{QCD}}^4} e^{-S},
\end{align}
where $\Lambda_{\mathrm{QCD}}\simeq 200~\mathrm{MeV}$ is the energy scale at which the QCD coupling becomes strong. We naively assume $c\sin\delta\sim\mathcal{O}(1)$. To satisfy the constraint imposed by neutron EDM measurements~\cite{Abel:2020pzs}, the value of the wormhole action must be sufficiently large, $S\gtrsim 2\times 10^2$, for typical values of $f_a$~\cite{Kallosh:1995hi,Alvey:2020nyh}.

\begin{figure}[t]
    \centering
    {\includegraphics[width=0.45\textwidth]{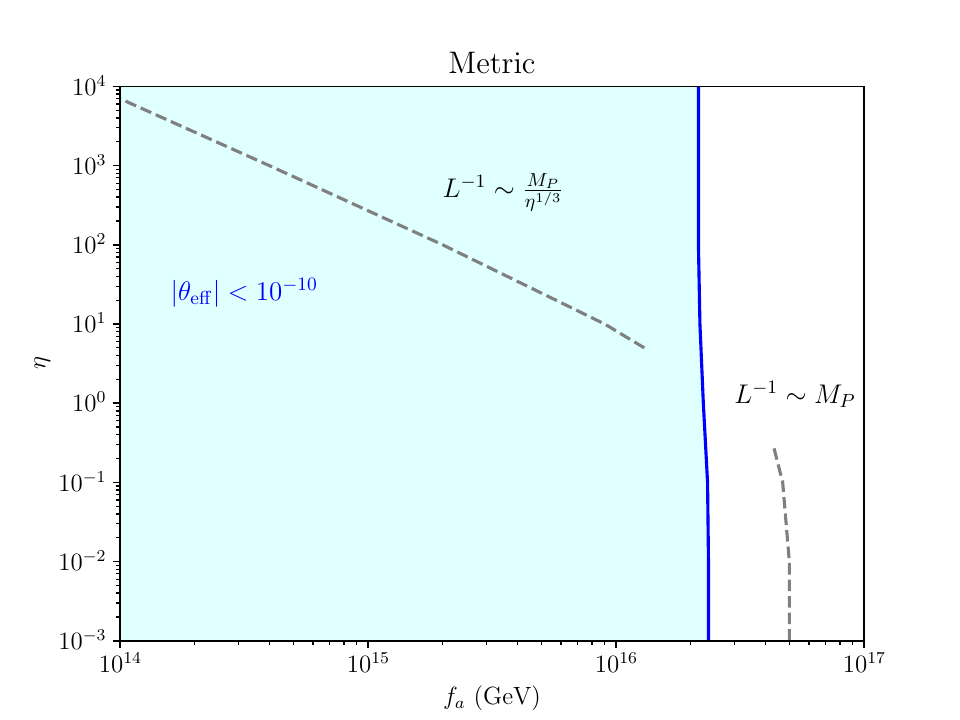}}
    {\includegraphics[width=0.45\textwidth]{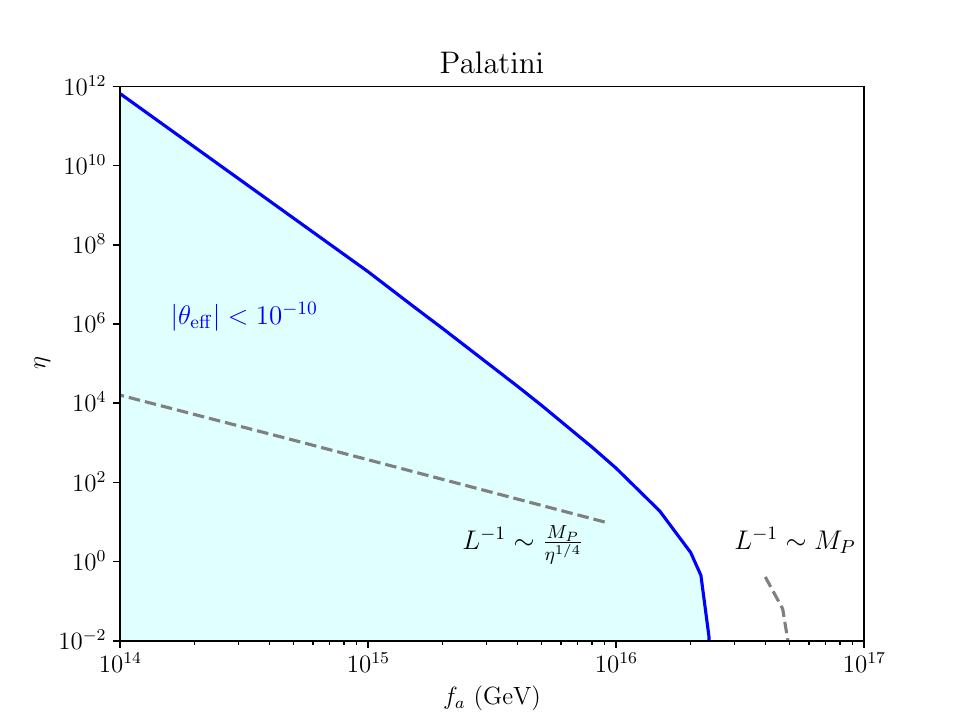}}
    \caption{$f_a-\eta$ plane in the metric and Palatini formulations. We assume $|c\sin\delta|=1$. Below the blue lines, the quality problem can be avoided. The dashed-gray lines show the contour of $L^{-1} \sim \Lambda$.}
    \label{fig:quality}
\end{figure}

In the metric formulation, the wormhole action remains at values around $\sim\frac{|n| M_P}{f_a}$ for any value of $\eta$. When the decay constant is $f_a\lesssim 2\times10^{16}~\mathrm{GeV}$, the wormhole-induced operators are adequately suppressed, even for unit PQ charges, as depicted in Fig.~\ref{fig:quality} (upper). In the Palatini formulation, the wormhole action is suppressed for large $\eta$, and the fine-tuning issue can be circumvented in the blue region in Fig.~\ref{fig:quality} (bottom). In both cases, the region above the dashed-gray line is excluded because $L^{-1}$ is larger than the cutoff scale $\Lambda$.

It is important to note that the axionic wormhole is one potential source of explicit PQ violation. Our analysis does not conclusively resolve the quality problem. Some arguments suggest the existence of non-perturbative effects that could explicitly violate the U(1) PQ symmetry~\cite{Dine:1986zy,Becker:1995kb,Gibbons:1995vg,Svrcek:2006yi,Alvey:2020nyh}.

\section{Conclusions and discussion}
\label{sec:conc}

We have analyzed axionic wormholes in models incorporating the $R^2$ term, considering both metric and Palatini formulations. It is essential to note that the Weyl transformation of the Ricci scalar significantly differs between these two formulations, and this discrepancy has a substantial impact on the characteristics of the wormhole solutions.
In the metric case, the Einstein frame action introduces an additional degree of freedom, denoted as $\omega$. The potential is minimized when $\omega=0$, resulting in an asymptotically flat spacetime.
On the other hand, in the Palatini formulation, the Einstein frame action lacks a kinetic term for the $\phi$ field. Consequently, the equation of motion yield a constraint on the fields. Utilizing this constraint, we establish the equivalence of the $R^2$ model to models involving higher-order derivatives.
In both formulations, we have demonstrated the existence of wormhole solutions for any positive value of $\eta$. However, it is worth noting that the existence of wormhole solutions is not always guaranteed. For instance, Ref.~\cite{Kallosh:1995hi} presents an example of the absence of wormhole solutions in a model with higher-order gravitational corrections.

We have computed the wormhole action and assessed the extent of explicit PQ violation in the $R^2$ models. In the metric formulations, the wormhole action remains sufficiently large to satisfy the neutron EDM constraint for $f_a\lesssim 2 \times 10^{16}~\mathrm{GeV}$, and this condition shows minor dependence on $\eta$. The presence of the quadratic scalar curvature results in only small alterations to the wormhole action, even for large values of $\eta$, when compared to the action in the Giddings-Strominger model.
Conversely, in the Palatini formulation, the wormhole action experiences suppression for large values of $\eta$. The axion quality problem can be mitigated as long as $\eta$ remains below an upper bound that depends on $f_a$, as illustrated in Fig.~\ref{fig:quality}. In the both cases, it is imperative to verify that the inverse size of the wormhole throat does not surpass the cutoff scale.

In our analysis, wormholes are supposed to contribute to the Euclidean path integral as saddle-point solutions. If these solutions are unstable, they could introduce non-trivial contributions to the quantum transition amplitudes. The stability of Giddings-Strominger wormholes has been a subject of debate for a long time~\cite{Rubakov:1996cn,Rubakov:1996br,Kim:1997dm,Hertog:2018kbz,Loges:2022nuw}. A recent stability analysis in the dual picture, described by a 3-form field, suggests that Giddings-Strominger wormholes do not possess negative modes, which correspond to lower action values~\cite{Loges:2022nuw}. A similar analysis may shed light on the stability of wormhole solutions in our setup.
\\

{\it Note added:}
During the completion of this manuscript, the paper~\cite{Cheong:2023hrj} has been submitted to the arXiv, which studies the axion wormholes using the effective field theory approach.
This paper explores various models, including those with the $R^2$ term, the dynamical radial mode of the PQ scalar field, and its non-minimal gravitational coupling.
While our analysis in the metric formulation corresponds to their induced gravity limit, our work in the Palatini formulation presents a novel and complementary contribution.

\section*{Acknowledgments}
This work is supported in part by the Grant-in-Aid for JSPS Fellows (No.21J20445 [YK]). The author thanks Koichi Hamaguchi and Natsumi Nagata for careful reading of the draft, useful discussions and comments.


\renewcommand{\theequation}{\Alph{section}.\arabic{equation}~}
\begin{appendix}


\end{appendix}

\bibliography{ref}

\end{document}